# Large Kerr nonlinearities on cavity-atom polaritons

Yifu Zhu

Department of Physics, Florida International University, Miami, FL

## Abstract

I analyze a scheme that is capable of producing large Kerr nonlinearities on cavity-atom polaritons in a cavity QED system consisting of multiple three-level atoms confined in a cavity mode. A weak control laser coupled to the atoms from free space induces destructive quantum interference in the polariton excitation of the coupled cavity-atom system and creates large Kerr nonlinearities on the intra-cavity light field. The scheme can be used for optical switching or cross-phase modulation of the cavity-atom polariton at ultra-low light levels, which may be useful for applications in quantum state manipulation in cavity QED, quantum measurements, and quantum logic gates.

Cavity quantum electrodynamics (Cavity QED) has a variety of applications in quantum physics and quantum electronics and has been a subject of many recent studies [1]. The fundamental cavity QED system consists of a single two-level atom coupled to a single cavity mode [2-3]. The two first-excited eigenstates of the coupled cavity-atom system are called the normal modes and are separated in energy by 2g, commonly referred to as the vacuum Rabi splitting ($g = \mu\sqrt{\omega_c/2\hbar\varepsilon_0 V}$ is the atom-cavity coupling coefficient, where μ is the atomic dipole moment, $\omega_c$ is the frequency of the cavity mode, V is the mode volume, $\hbar$ is the Plank constant, and $\varepsilon_0$ is the vacuum permittivity). The two normal modes consist of coherent superposition of the atomic excitation and photonic excitation, and equivalently can be viewed as polaritons of the coupled cavity-atom system. If N two-level atoms interact collectively with the cavity mode, the coupling coefficient becomes $G = \sqrt{N}g$ and the vacuum Rabi splitting of the normal modes for the collectively coupled atom-cavity system becomes 2G and may be observed in a cavity with a moderate mode volume and finesse [4-6].

Here we propose a scheme for producing large Kerr nonlinearities and cross-phase modulation (XPM) for the cavity-atom polaritons in a multi-atom cavity QED system. The cavity-atom polaritons are excited by coupling a probe laser into the cavity mode containing N three-level atoms. We show that with a weak control laser coupled to the atoms from free space and tuned in frequency to near the polariton (the normal mode) resonance, destructive quantum interference is induced in the polariton excitation, which results in large $\chi^{(3)}$ nonlinearities for the intra-cavity light field. On resonance, the polariton excitation is suppressed by the imaginary part of $\chi^{(3)}$ representing the nonlinear absorption; near the resonance, the real part of $\chi^{(3)}$, the Kerr nonlinearity, produces large XPM on the excited polaritons, which can be measured via the phase shift of the cavity transmitted light.

Third-order susceptibilities $\chi^{(3)}$ in an optical medium have many applications in nonlinear optics [7]. Recent studies have shown that $\chi^{(3)}$ plays an important role for quantum nondemolition measurements, quantum information processing, generation of single photons and correlated photon pair, and nonlinear light control [8-11]. In those applications, it generally requires large values of $\chi^{(3)}$ under conditions of low light powers and high sensitivities. Several schemes for producing large $\chi^{(3)}$ susceptibilities at weak light intensities have been analyzed and explored [12-18]. The proposed scheme here is capable of producing large Kerr nonlinearities on the cavity-atom polaritons with a control field at ultra-low intensities. It combines fundamental studies of cavity QED and nonlinear optics at ultra-low light levels, and should be useful for applications such as quantum state manipulation in cavity QED, quantum measurements, and quantum phase gates.

The proposed scheme is shown in Fig. 1. The composite atom-cavity system consists of a single mode cavity containing N identical three-level atoms [19-22]. The cavity consists of two mirrors with identical reflectivity R and separated by a length L. The cavity mode couples the atomic transition |a>-|e> and is tuned to the atomic transition frequency $\nu_{ea}$ ($\Delta_c = \nu_c - \nu_{ea} = 0$). A weak probe (signal) laser is coupled into the cavity mode and at appropriate frequency, excites the cavity-atom polaritons. A weak control laser couples the atoms from open free space and drives the atomic transition |b>-|e> with Rabi frequency $2\Omega$. $\Delta = \nu - \nu_{eb}$ is the control frequency detuning, and $\Delta_p = \nu_p - \nu_{ea}$ is the probe frequency detuning. The collective atomic operators are $\rho_{ae} = \sum_{i=1}^{N} \rho_{ae}^i$ and $\rho_{be} = \sum_{i=1}^{N} \rho_{be}^i$, and couple the symmetric, Dicke-type atomic states, among which the ground state is represented by $|a>=|a_1.........a_N>$ and the excited state with one atomic excitation is represented by $|e\rangle = \sum_{j=1}^{N} \frac{1}{\sqrt{N}} |a_1......e_j......a_N\rangle$. Without the weak control laser, the ground state of the cavity-atom system is $|a,0\rangle = |a\rangle|0\rangle$, and the two product states with one excitation quanta are $|a,1\rangle = |a\rangle|1\rangle$ and $|e,0\rangle = |e\rangle|0\rangle$ (|1> and |0> are one-photon and zero photon states of the cavity mode). The coupling of the cavity mode and the collective atomic states produces multi-ladder eigenstates, the lowest three of which are the unchanged ground state $|a,0\rangle = |a\rangle|0\rangle$, and first excited states, $|\varphi_+\rangle = \frac{1}{\sqrt{2}}(|e,0\rangle + |a,1\rangle)$ and $|\varphi_-\rangle = \frac{1}{\sqrt{2}}(|e,0\rangle - |a,1\rangle)$. The two excited states are separated in frequency by the vacuum Rabi splitting $2\sqrt{N}g$ and are commonly referred to as two normal modes of the coupled cavity-atoms system, or equivalently, polaritons of the atomic excitation and photonic excitation [4-6]. The linewidth of the two normal modes is given by $(\kappa + \Gamma)/2$, the average of the cavity linewidth $\kappa = c(1-R)/(2\pi L \sqrt{R})$ (c is the light speed in vacuum) and the atomic linewidth $\Gamma$ [5]. When the weak control field with $\Omega \ll \sqrt{N}g$ is present and tuned to be resonant with either one of the two normal modes ($\Delta = \sqrt{N}g$ or $-\sqrt{N}g$), it creates two polariton excitation paths and the resulting destructive quantum interference between the two paths suppresses the polartion (normal mode) excitation [23]. Here we analyze the nonlinear susceptibilities induced by the control laser near the polariton resonance and show that large Kerr nonlinearities can be created in the cavity-atom system. The Kerr nonlinearities modify the polariton amplitude and result in a large XPM phase shift for the excited polariton, which can be measured from the phase shift and the intensity of the cavity transmitted probe field.

We treat the cavity field classically and derive the atomic susceptibilities in a semiclassical analysis in which the atoms are coupled by a free space control laser and a weak, intra-cavity probe laser. The susceptibility of the three-level Λ-type atomic medium at the frequency of the intra-cavity probe laser $\nu_p$, $\chi(\upsilon_p) = \chi'(\upsilon_p) + i\chi''(\upsilon_p)$, is derived by solving Schrodinger equations of the coupled atomic system under the condition $\rho_{aa} \approx 1$ (the intra-cavity probe field is much weaker than the coupling field so the atomic population is concentrated in the ground state |a>) and is given by

$$\chi(\nu_p) = \chi' + i\chi'' = \frac{K(\Delta_p - \Delta + i\gamma_{ab})}{|\Omega|^2 - (\Delta_p + i\Gamma/2)(\Delta - \Delta_p + i\gamma_{ab})}. \qquad (1)$$

Here $K = n|\mu_{ea}|^2/\hbar\varepsilon_0$ (n is the atomic density), $\Gamma$ is the decay rate of the excited state |e> and $\gamma_{ab}$ is the decoherence rate between ground states |a> and |b>. The real part of the susceptibility, $\chi'$ contributes to the phase shift of the intra-cavity probe field while the imaginary part, $\chi''$, results in the attenuation of the intra-cavity probe field. The amplitude of the cavity-transmitted probe field is

$$E_t(\upsilon_p) = |E_t|\exp(i\varphi_t) = \frac{E_{in}(\nu_p)(1-R)\exp(ik(L - \ell + \chi'\ell + i\chi''\ell))}{(1 - R \cdot \exp(2ik(L - \ell + \chi'\ell + i\chi''\ell)))} \qquad (2).$$

and the phase shift is given by

$$\Phi_t = \tan^{-1}\{\frac{1 + R \cdot \exp(-2k\chi''\ell)}{1 - R \cdot \exp(-2k\chi''\ell)} \tan(k(L - \ell + \chi'\ell))\}, \qquad (3)$$

Here $\ell$ is the atomic medium length in the cavity mode. With the resonant frequency of the empty cavity locked to the atomic transition frequency ($\Delta_c = \nu_c - \nu_{ea} = 0$), we calculate the transmitted probe field for a practical cavity-atom system (L=5 cm, R=0.98, $\ell$=1 mm, and a realistic optical depth of the cold Rb atoms, OD=$n\sigma_{ea}\ell \sim 2 - 8$). If the control laser is absent, the susceptibilities are linear and given by $\chi = \frac{K(i\Gamma/2 - \Delta_p)}{\Delta_p^2 + (\Gamma/2)^2}$. Fig. 2(a) plots the transmitted probe intensity $I_t$ normalized to the input probe intensity $I_{in}$ versus the normalized probe frequency detuning $\Delta_p/\Gamma$ and exhibits two transmission peaks located at $\Delta_p = \pm\sqrt{\frac{k\ell cK}{2\pi L} - \frac{\Gamma^2}{4}} = \pm\sqrt{\frac{|\mu_{ea}|^2 v_p n\ell}{2\hbar\varepsilon_0 L} - \frac{\Gamma^2}{4}}$ as a result of cancelation of the probe phase shift from the atoms, $k\chi'\ell$, and the probe phase shift from the empty cavity, $k(L-\ell)$ [24]. The number of atoms inside the intra-cavity probe beam is $N=nA\ell=ODA/\sigma_{ea}$ (A is the cross section of the intra-cavity probe beam) and the cavity mode volume is $V=AL$. Then the two transmission peaks occur at $\Delta_p = \pm\sqrt{\frac{|\mu_{ea}|^2 v_p n\ell}{2\hbar\varepsilon_0 L} - \frac{\Gamma^2}{4}} = \pm\sqrt{\frac{|\mu_{ea}|^2 v_p N}{2\hbar\varepsilon_0 V} - \frac{\Gamma^2}{4}} = \pm\sqrt{Ng^2 - \frac{\Gamma^2}{4}}$. When $4Ng^2 \gg \Gamma^2$, $\Delta_p = \pm\sqrt{N}g$. In Cavity QED, the multi-atom vacuum Rabi splitting (the normal mode splitting) is given by $2\sqrt{N}g$ (with $4Ng^2 \gg \Gamma^2$) [4-6]. Thus, the semiclassical analysis gives the identical result for the vacuum Rabi splitting as the QED analysis and the two transmission peaks at $\Delta_p=\pm\sqrt{N}g$ correspond to the two polariton resonances, or the two excited normal modes.

When the control laser is present, the atomic coherence and interference is induced in the coupled cavity-atom system. It is instructive to exam the dependence of $\chi$ on the control laser detuning $\Delta$ near the polariton resonance when the probe laser frequency is locked to the polariton resonance at $\Delta_p=\sqrt{N}g$ (or $\Delta_p=-\sqrt{N}g$). Under the condition $\gamma_{ab}=0$ (neglecting the ground state decoherence), when $\Delta_p-\Delta=0$ (the control laser is exactly on the polariton resonance $\Delta=\Delta_p=\sqrt{N}g$ (or $-\sqrt{N}g$), $\chi=0$, the susceptibilities vanishes and the coupled cavity-atom system behaves like an empty cavity (with only the phase shift from the empty cavity, $k(L-\ell)$). Since the empty-cavity resonance is set at $\Delta_c = v_c - v_{ea}=0$ and the probe laser frequency is detuned from the empty-cavity resonance by $\sqrt{N}g$ (or $-\sqrt{N}g$), the probe laser cannot be coupled into the cavity, i.e., the excitation of the cavity-atom polaritons is suppressed. Fig. 2(b) plots the transmitted probe field versus $\Delta_p/\Gamma$ when the weak control laser is present and tuned to $\Delta=\sqrt{N}g$. It shows that a narrow dip appears at the transmission peak $\Delta_p=\sqrt{N}g$, representing suppression of the polariton excitation at $\Delta_p=\sqrt{N}g$. The linewidth of the dark dip is significantly narrower than the natural linewidth $\Gamma$ and the empty cavity linewidth $\kappa$, a signature of quantum coherence and interference induced by the control laser [23].

When $\Delta-\Delta_p\neq 0$, the control laser is tuned away from the polariton resonance, the Kerr nonlinearity is generated and results in a large XMP of the excited polariton, which can be measured from the phase shift of the transmitted probe field. With the probe frequency locked to the polariton resonance at $\Delta_p=\sqrt{N}g$ (or $-\sqrt{N}g$) and writing $\Delta=\Delta_p+\delta$ ($\delta$ is the control frequency detuning from the polariton resonance), Eq.(1) is reduced to (for a weak control field satisfying $|\Omega|^2 \ll |\sqrt{N}g\delta|$)

$$\chi' = \chi'^{(1)} + \chi'^{(3)}|E|^2 = -\frac{K\Delta_p}{\Delta_p^2 + (\Gamma/2)^2} + \frac{K|\mu_{eb}|^2 (\Delta_p^2 - (\Gamma/2)^2)}{\hbar^2 (\Delta_p^2 + (\Gamma/2)^2)^2 \delta}|E|^2, \quad (4)$$

$$\chi'' = \chi''^{(1)} + \chi''^{(3)}|E|^2 = \frac{K\Gamma/2}{\Delta_p^2 + (\Gamma/2)^2} - \frac{K|\mu_{eb}|^2 \Delta_p \Gamma}{\hbar^2 (\Delta_p^2 + (\Gamma/2)^2)^2 \delta}|E|^2. \quad (5)$$

Here the linear susceptibilities are given by $\chi^{(1)}=\chi'^{(1)}+i\chi''^{(1)}$ and the third order susceptibilities induced by the control field E are given by $\chi^{(3)}=\chi'^{(3)}+i\chi''^{(3)}$. As discussed before, at $\Delta_p=\sqrt{N}g$ (or $-\sqrt{N}g$), the phase shift of the linear dispersion, $k\chi'^{(1)}\ell$, cancels the empty-cavity phase shift, $k(L-\ell)$, which results in the polariton (normal mode) excitation (the probe transmission peak at $\Delta_p=\sqrt{N}g$ or $-\sqrt{N}g$) [24]. Then, the phase shift of the intra-cavity probe field is solely from the control-field induced Kerr nonlinearity $\chi'^{(3)}$. The imaginary part of third-order nonlinearities, $\chi''^{(3)}$, attenuates the amplitude of the intra-cavity field. That is, the third-order nonlinearities induced by the control laser change the polariton phase and amplitude. With an appropriate δ value, the polariton experiences a large XPM phase shift but can still keep sufficiently large amplitude. When $\Delta_p = \sqrt{N}g \gg \Gamma$ and $\sqrt{N}g \gg |\Omega^2/\delta|$, one derives $\chi'^{(3)} = \frac{K|\mu_{eb}|^2}{\hbar^2 g^2 N \delta}$ and $\chi''^{(3)} = -\frac{K|\mu_{eb}|^2 \Gamma}{\hbar^2 g^3 N^{3/2} \delta}$. Since $K = n|\mu_{ea}|^2/\hbar\varepsilon_0$ and $N = nA\ell$, the Kerr nonlinearity $\chi'^{(3)}$ is independent of n, the atomic number density in the cavity mode, and consequently, the induced XPM phase shift on the cavity-atom polariton is independent of N, the total number of atoms in the cavity mode. This seemingly unexpected result comes from the fact that the XPM phase shift is proportional to $N/\Delta_p^2$ in which $\Delta_p = \sqrt{N}g$ at the polariton resonance. The nonlinear absorption of the intra-cavity probe light is given by $\chi''^{(3)}$ which is then proportional to $1/\sqrt{N}$. Under these conditions, the single path XPM phase shift is $k\chi'^{(3)}|E|^2 \ell$ and the nonlinear absorption length is $2k\chi''^{(3)}|E|^2 \ell$. With the feedback enhancement of multiple reflections in the cavity, the XPM phase shift becomes $2Fk\chi'^{(3)}|E|^2 \ell/\pi$ ($F = \pi\sqrt{R}/(1-R)$ is the cavity finesse). For a cavity-atom system with $\Omega=0.2\Gamma$, $R=0.98$, $n\sigma_{ea}\ell=4$, and $\delta=0.1\Gamma$, the total phase shift is $\Phi_t \approx 0.4$ rad. and the estimated transmitted light intensity is 92% of the transmitted probe light without the control laser. Larger phase shifts can be obtained for a cavity with higher finesse values (with R>0.98) and/or smaller δ values (with the condition $|\Omega|^2 \ll |\sqrt{N}g\delta|$ satisfied). The figure of merit for the Kerr nonlinearities is defined as $\eta = \frac{\chi'^{(3)}}{\chi''^{(3)}}$ [12-13]. For the proposed scheme here under the conditions of $\sqrt{N}g \gg \Gamma$ and $\sqrt{N}g \gg |\Omega^2/\delta|$, $\eta = \frac{\chi'^{(3)}}{\chi''^{(3)}} \approx \frac{\Delta_p}{\Gamma} = \frac{g\sqrt{N}}{\Gamma} \gg 1$.

To quantify above discussions, we present numerical calculations from equations (1) and (2) in Fig. (3) and Fig. (4) for the cold Rb atoms and cavity system used in Fig. 2. Fig. 3 plots (a) the phase shift $\Phi_t$ and (b) the transmitted intensity of the probe field versus the control detuning $\delta$, which shows the spectral profile of the pure $\chi^{(3)}$ contribution. In order to reveal the effect of the decoherence rate $\gamma_{ab}$ on the Kerr nonlinearities, a set of curves with different $\gamma_{ab}$ values are plotted together. Fig. 3 shows that the decoherence reduces the XPM phase shift and the interference dip (giving rise to a larger transmission intensity). The interference dip of the polariton excitation (the cavity transmission dip) exhibits a near Lorentzian line profile with a linewidth ultimately limited by the decoherence rate $\gamma_{ab}$. With $\Omega<<\Gamma$, the linewidth is significantly narrower than $\Gamma$.

We note that the lifetime in the Rb excited state is ~ 30 ns, and the life time of the ground state coherence as long as a few ms has been observed in the experiments with cold Rb atoms [25], which corresponds to $\gamma_{ab}$~$10^{-4}\Gamma$. In order to see the orders of magnitude of the Kerr nonlinearities obtainable from experiments with cold Rb atoms, we extract some numerical values from Fig. 3 (with the decoherence rate $\gamma_{ab}$=0.01$\Gamma$, a conservative estimate). Without the control laser, the transmitted probe intensity is $I_t/I_{in}$=58% (at the full polariton amplitude); with a weak control laser ($\Omega$=0.2$\Gamma$), the calculated phase shift is $\Phi_t$≈0.61 rad. and the transmitted probe intensity $I_t/I_{in}$≈48% (83% of the full polariton intensity) are obtained for $\delta$=0.05$\Gamma$; if $\delta$=0.1$\Gamma$, the phase shift $\Phi_t$≈0.35 rad. and the transmitted probe intensity $I_t/I_{in}$≈56% (93% of the full polariton intensity) are obtained. These results are obtained under the condition $\sqrt{N}g>>\Gamma$ and $\sqrt{N}g>>|\Omega_c^2/\delta|$ and are consistent with the analytical results presented before, Therefore, large XPM phase shifts can be obtained while sufficiently large amplitudes of the excited polaritons are preserved. The calculations are obtained with the control Rabi frequency $\Omega=0.2\Gamma$, well below the saturation intensity. For comparison, we note that for the $\Lambda$-type system formed by Rb D2 transitions, a single photon (780 nm) in a 1 µs pulse and focused to a spot size of a half wavelength has an intensity of 0.17 mW/cm$^2$, which corresponds to $\Omega=0.35\Gamma$ (the Rb saturation intensity is 1.6 mW/cm$^2$). Therefore, the proposed system is capable of performing the large XPM of the cavity-atom polaritons at ultra-low levels of the control light field.

Fig. 4 plots the phase shift (a) and transmitted probe light versus the control detuning $\Delta/\Gamma$ for several values of OD=$n\sigma_{ea}\ell$. The line shape, and the amplitude of the XPM phase shift, and the amplitude of the transmitted probe intensity are nearly identical for curves 2-5 in Fig. 4, which confirms that when $\sqrt{N}g\,|\delta|>>|\Omega|^2$ and $\sqrt{N}g>>\Gamma$, the Kerr nonlinearities induced by the control laser are independent of n, atomic number density and the XPM phase shift is independent of N. When the condition, $\sqrt{N}g>>\Gamma$, is invalid, the Kerr nonlinearity decreases for smaller n values as shown by curve 1 in Fig. 4.

To observe the effects of the control field amplitude on the induced Kerr nonlinearities, we plot in Fig. 5 the phase shift $\Phi_t$ (Fig 5(a)) and the transmitted intensity of the probe field (Fig. 5(b)) versus the control Rabi frequency $\Omega$ as well as the control detuning $\delta$. As the control field intensity increases, the XPM phase shift increases and the transmitted light intensity decreases. The spectral linewidth of the interference dip is broadened at larger $\Omega$ values, showing the characteristic of the power broadening effect.

In conclusion, we have proposed and analyzed a scheme for generating large Kerr nonlinearities with a weak control laser in a coupled cavity-atom system. The scheme can be used for optical switching of the cavity-atom polaritons and for obtaining a large XPM phase shift of the excited polaritons by an ultra-weak control laser. The results show that the large XPM on the cavity-atom polaritons can be obtained in a practical experimental system with cold

Rb atoms confined in a moderate finesse cavity. The analysis shows that if $\sqrt{N}g \gg \Gamma$ and $\sqrt{N}g \gg |\Omega^2/\delta|$, the XPM phase shift on the excited polariton is independent of N. The scheme is expected to be applicable to the single-atom cavity QED system where the strong coupling gives rise to a vacuum Rabi frequency satisfying the condition g$\gg\Gamma$ and $g \gg |\Omega^2/\delta|$ (for N=1) and also may be useful for semiconductor cavity QED systems [26]. It will be interesting to explore possible applications of the scheme in quantum phase gates and quantum measurements.

This paper is based upon work supported by the National Science Foundation under Grant No. 0757984.

References

1. *Cavity Quantum Electrodynamics*, edited by P. R. berman (Academic, San Diego, 1994).
2. A. Boca, R. Miller, K. M. Birnbaum, A. D. Boozer, J. McKeever, and H. J. Kimble, "Observation of the vacuum Rabi spectrum for one trapped atom", Phys. Rev. Lett. **93**, 233603 (2004).
3. T. Puppe, I. Schuster, A. Grothe, A. Kubanek, K. Murr K, P. Pinkse, and G. Rempe, "Trapping and observing single atoms in a blue-detuned intracavity dipole trap", Phys. Rev. Lett. **99**, 013002( 2007).
4. G. S. Agarwal, "Vacuum-Field Rabi Splittings in Microwave Absorption by Rydberg Atoms in a Cavity", Phys. Rev. Lett. **53**, 1732(1984).
5. M. G. Raizen, R. J. Thompson, R. J. Brecha, H. J. Kimble, and H. J. Carmichael, "Normal-mode splitting and linewidth averaging for two-state atoms in an optical cavity", Phys. Rev. Lett. **63**, 240 - 243 (1989).
6. Y. Zhu, D. J. Gauthier, S. E. Morin, Q. Wu, H. J. Carmichael, and T. W. Mossberg, "Vacuum Rabi splitting as a feature of linear-dispersion theory: Analysis and experimental observations", Phys. Rev. Lett. **64**, 2499 (1990).
7. Y. R. Shen, *The Principles of Nonlinear Optics*, Wiley, New York, 1984.
8. S. E. Harris and L. V. Hau, "Nonlinear optics at low light levels", Phys. Rev. Lett. **82**, 4611 (1999).
9. M. D. Lukin, and A. Imamoglu, "Nonlinear Optics and Quantum Entanglement of Ultraslow Single Photons", Phys. Rev. Lett. **84**, 1419 (2000).
10. D. A. Balic , V. Balicm, S. Goda, G. Y. Yin, and S. E. Harris, "Frequency mixing using electromagnetically induced transparency in cold atoms", Phys. Rev. Lett. **93**, 183601 (2004).
11. D. Vitali, M. Fortunato, P. Tombesi, "Complete quantum teleportation with a Kerr nonlinearity", Phys. Rev. Lett. **85**, 445 (2000).
12. H. Schmidt and A. Imamoglu, "Giant Kerr nonlinearities obtained by electromagnetically induced transparency", Opt. Lett. **21**, 1936 (1996).
13. H. Kang and Y. Zhu, "Observation of large Kerr nonlinearity at low light intensities", Phys. Rev. Lett. **91**, 93601 (2003).
14. H. Wang, D. Goorskey, M. Xiao, "Enhanced Kerr nonlinearity via atomic coherence in a three-level atomic system", Phys. Rev. Lett. **87**, 073601(2001).
15. A. B. Matsko, I. Novikova, G. R. Welch, and M. S. Zubairy, "Enhancement of Kerr nonlinearity by multiphoton coherence", Opt. Lett. **28**, 96 (2003).
16. D. Petrosyan and Y. P. Malakyan, "Magneto-optical rotation and cross-phase modulation via coherently driven four-level atoms in a tripod configuration", Phys. Rev. A **70**, 023822(2004).
17. Y. F. Chen, C. Y. Wang, S. H. Wang, and I. A. Yu, "Low-Light-Level Cross-Phase Modulation Based on Stored Light Pulses", Phys. Rev. Lett. **96**, 043603 (2006).
18. Z. B. Wang, K. P. Marzlin, B. C. Sanders, "Large cross-phase modulation between slow copropagating weak pulses in Rb-87", Phys. Rev. Lett. **97**, 063901(2006).
19. M. D. Lukin, M. Fleischhauer, M. O. Scully, and V. L. Velichansky, "Intracavity electromagnetically induced transparency ", Opt. Lett. **23**, 295 (1998).
20. H. Wang, D. J. Goorskey, W. H. Burkett, and M. Xiao, "Cavity-linewidth narrowing by means of electromagnetically induced transparency ", Opt. Lett. **25**, 1732 (2000).
21. G. Hernandez, J. Zhang, and Y. Zhu, "Vacuum Rabi splitting and intracavity dark state in a

cavity-atoms system", Phys. Rev. A **76**, 053814 (2007).
22. H. Wu, J. Gea-Banacloche, and M. Xiao, "Observation of Intracavity Electromagnetically InducedTransparency and Polariton Resonances in a Doppler-Broadened Medium," Phys. Rev. Lett. 100, 173602(1-4) (2008).
23. J. Zhang, G. Hernandez, and Y. Zhu, "Suppressing normal mode excitation by quantum interference in a cavity-atom system", Optics Express **16**, 7860(2008).
24. Gessler Hernandez, Jiepeng Zhang, and Yifu Zhu, "Collective coupling of atoms with cavity mode and free-space field", Optics Express **17**, 4798(2009).
25. R. Zhao, T. O. Oudin, S. D. Jenkins, C. J. Campbell, D. N. Matsukevich, T. A. B. Kenndedy, and A. Kuzmich, "Long-liver quantum memory", Nature Phys. **5**, 100(2009)
26. G. Khitrova, H. M. Gibbs, M. Kira, S. W. Koch, and A. Scherer, "Vacuum Rabi splitting in semiconductors", Nature Phys. **2**, 81(2006).

Figure Captions

Fig. 1 Three level atoms coupled to a cavity field and a free-space control field.

Fig. 2 $I_t / I_{in}$ ($I_t$ is the cavity transmitted probe intensity and $I_{in}$ is the input probe intensity) versus the normalized probe frequency detuning $\Delta_p/\Gamma$. (a) Without the control laser. (b) With the control laser ($\Delta=\sqrt{N}g$ and $\Omega=0.2\Gamma$). Parameters used in the calculations are: $\gamma= 0.01\Gamma$, optical depth $n\sigma_{ea}\ell=3$, L=5 cm, $\Delta_c=0$, and R=0.98. The resulting vacuum Rabi frequency $2\sqrt{N}g =19\Gamma$. The inset Fig. plots the intensity (top) and the phase shift (bottom) of the transmitted probe light between $\Delta_p/\Gamma= 8$ to 11.

Fig. 3 (a) Phase shift $\Phi_t$ and (b) normalized intensity $I_t / I_{in}$ of the transmitted probe field versus the normalized control frequency detuning $\delta/\Gamma$ ($\delta=\Delta - \sqrt{N}g$ ). Black line, $\gamma_{ab}= 0$; green line, $\gamma_{ab} = 0.001\Gamma$; blue line, $\gamma_{ab} = 0.01\Gamma$; red line, $\gamma_{ab} = 0.02\Gamma$, and dark green line, $\gamma_{ab} = 0.05\Gamma$. The optical depth $n\sigma_{ea}\ell=4$. The other parameters are the same as those in Fig. 2.

Fig. 4 (a) Phase shift $\Phi_t$ and (b) normalized intensity $I_t / I_{in}$ of the transmitted probe field versus the normalized control frequency detuning $\Delta/\Gamma$. The FWHM linewidth of curves 2-5 in (b) is ≈0.05Γ. Curve 1, $n\sigma_{ea}\ell=0.1$ ($\Delta_p=\sqrt{N}g =1.7\Gamma$); curve 2, $n\sigma_{ea}\ell=2$ ($\Delta_p=\sqrt{N}g =7.8\Gamma$); curve 3, $n\sigma_{ea}\ell=4$ ($\Delta_p=\sqrt{N}g =11\Gamma$); curve 4, $n\sigma_{ea}\ell=6$ ($\Delta_p=\sqrt{N}g =13.5\Gamma$); and curve 5, $n\sigma_{ea}\ell=8$ ($\Delta_p==\sqrt{N}g =15.5\Gamma$). The other parameters are the same as those in Fig. 2.

Fig. 5 Phase shift $\Phi_t$ and (b) normalized intensity $I_t / I_{in}$ of the transmitted probe field versus the normalized control frequency detuning $\delta/\Gamma$ and control Rabi frequency $\Omega/\Gamma$. $\gamma_{ab}= 0.01\Gamma$, optical depth $n\sigma_{ea}\ell=4$, and the other parameters are the same as those in Fig. 2.

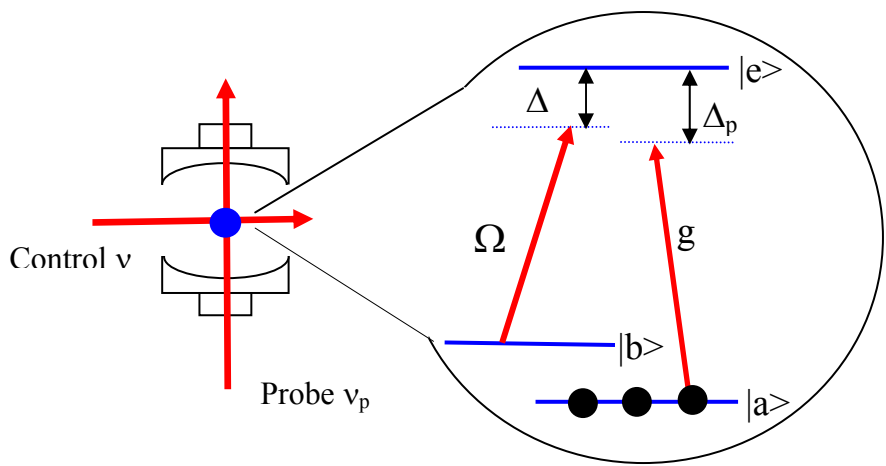

Fig. 1

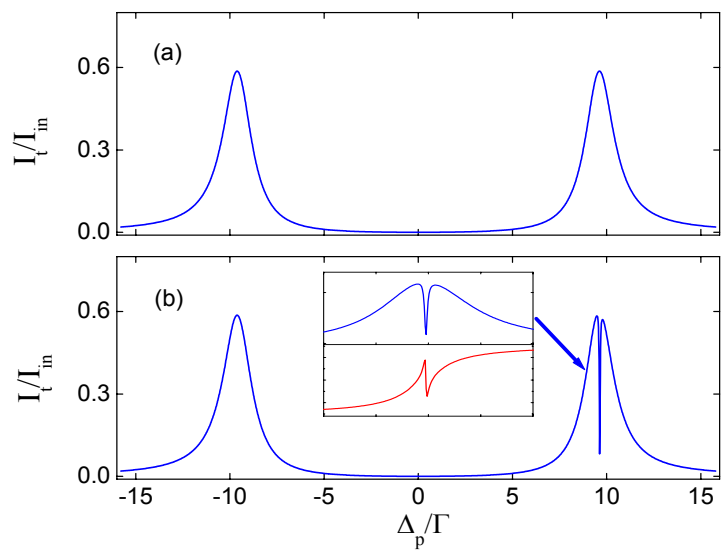

Fig. 2

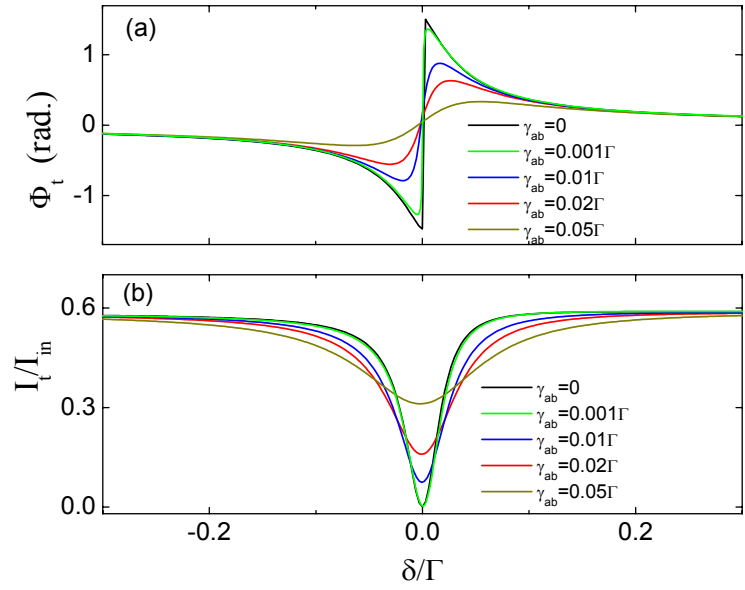

Fig. 3

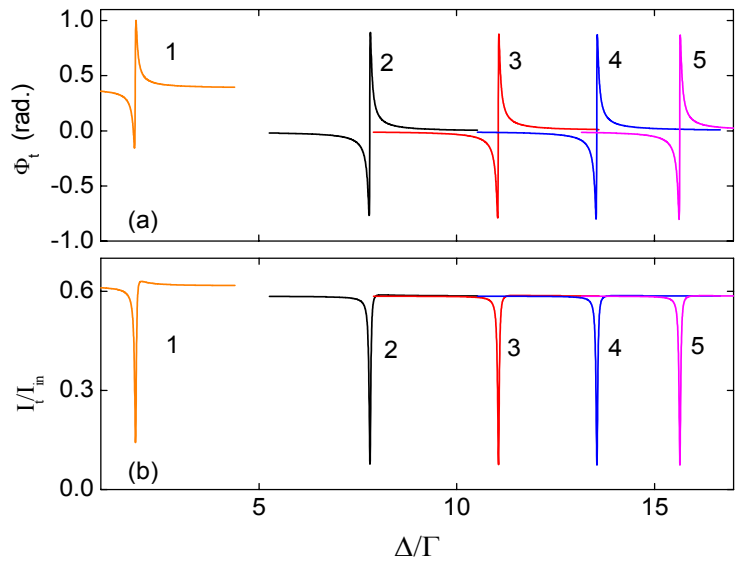

Fig. 4

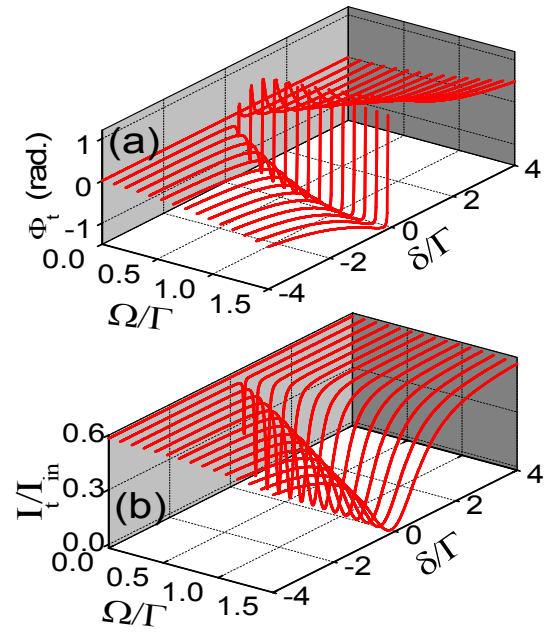

Fig. 5